\documentclass[english]{article}
\usepackage[T1]{fontenc}
\usepackage[latin9]{inputenc}
\usepackage{geometry}
\usepackage{array}
\usepackage{longtable}
\usepackage{graphicx}
\usepackage{cite,color}

\makeatletter

\providecommand{\tabularnewline}{\\}

\usepackage{babel}

\usepackage{babel}

\makeatother

\usepackage{babel}
\begin{document}
%\title{How secure is online banking: A systematic study of the security%provided by the major Indian banks\\% or\\

\title{How vulnerable are the Indian banks: A cryptographers' view}

\maketitle
\begin{center}
Anirban Pathak$^{\dagger,}$\footnote{email: anirban.pathak@gmail.com},
Rishi Dutt Sharma$^{\ddagger,}$\footnote{email: rishi.dtu@gmail.com}, Dhananjoy
Dey$^{\star,}$\footnote{email: dhananjoydey@sag.drdo.in}\\ 
$^{\dagger}$Jaypee Institute of Information Technology, A 10, Sector 62,
Noida, UP-201309, India\\
 $^{\ddagger}$Department of Computer Science Engineering, Bennett University,
Greater Noida, UP-201310, India\\
$^{\star}$Scientific Analysis Group, Metcalfe House Complex, Delhi-110
054, INDIA 
\par\end{center}

\begin{abstract}
With the advent of e-commerce and online banking it has become extremely
important that the websites of the financial institutes (especially,
banks) implement up-to-date measures of cyber security (in accordance
with the recommendations of the regulatory authority) and thus circumvent
the possibilities of financial frauds that may occur due to vulnerabilities
of the website. Here, we systematically investigate whether Indian
banks are following the above requirement. To perform the investigation,
recommendations of Reserve Bank of India (RBI), National Institute
of Standards and Technology (NIST), European Union Agency for Network
and Information Security (ENISA) and Internet Engineering Task Force
(IETF) are considered as the benchmarks. Further, the validity and quality
of the security certificates of various Indian banks have been tested
with the help of a set of tools (e.g., SSL Certificate Checker provided
by Digicert and SSL server test provided by SSL Labs). The analysis
performed by using these tools and a comparison with the benchmarks,
have revealed that the security measures taken by a set of Indian
banks are not up-to-date and are vulnerable under some known attacks. 
\end{abstract}

\section{Introduction}

Indian economy is one of the largest and fastest growing economy of
the world, and with the recent initiatives of the government, it is
moving towards complete digitization. In a digital economy, secrecy
of the exchanged information would play a crucial role. In India,
several steps have been taken by the banks and government agencies
for the orientation of the end users and thus to reduce the risk of
simple frauds that originate due to the mistakes of the users. However,
there is another facet of the security of e-commerce, which involves
the service providers and requires that the banks provide highest
possible security without becoming slow. To ensure that the Reserve
Bank of India (RBI), which regulates Indian banks have issued several
circulars \cite{RBI-1,RBI-2}, and have advised Indian banks to use
up-to-date encryption techniques \cite{RBI-1,RBI-2,moondra}. Here, we aim to check, to what extend
the Indian banks follow the advices issued by the RBI and in what
cases the security measures taken by the banks are vulnerable. We
would also suggest some solutions for the safe operations of the e-banking
portals in the present and future. This investigation would remain
focused on the Indian banks, and would aim to reveal such cases where
the security measures taken by the Indian banks are not up-to-date
and are vulnerable under some known attacks. To reveal these vulnerabilities,
we will adopt a more technical approach compared to the earlier studies
on the security of Indian banks and customer perspectives about that
\cite{Tejinder,Bhutt-2011-India, Gupta07}. Specifically, the validity and
quality of the security certificates of various Indian banks will
be tested here with the help of a set of tools (e.g., SSL Certificate
Checker provided by Digicert and SSL server test provided by SSL Labs).
The analysis performed by using these tools and a comparison with
the benchmarks, have revealed that the encryption techniques used
by many Indian banks are vulnerable. The uniqueness of the present
study is that to the best of our knowledge, no such technical analysis
of the encryption mechanism adopted by the Indian banks has been done
until now.

We have already mentioned that the security and privacy issues of
Indian banks have been studied earlier \cite{Tejinder,Bhutt-2011-India, Gupta07},
but a less technical view was adopted. Specifically, in Ref \cite{Tejinder},
possible attacks on customers' information like phishing, spoofing,
vishing had been mentioned. Some efforts had also been made to discuss
card-related frauds, merchant-related frauds, but no efforts had been
made to investigate the vulnerability of the banks. In another study
in the similar line (but not restricted to Indian banks) \cite{Hamed2017},
various dimensions of e-service security had been listed and utilized
to obtain a hierarchical structure of them to compute the weights
of security dimensions. Once again, the analysis was more focused
on the perspectives of the end users and led a trivial conclusion
that ``users will intend to use e-service if they feel that the quality
of e-service is high''. The same conclusion regarding the behavior
of end users has recently been reported again by Taherdoost in \cite{Hameed2018}
through E-Service Technology Acceptance Model (ETAM) which can evaluate
the user acceptance of e-service technology even before the introduction
of the service. Such studies on the acceptability of an e-service
by the end users is not new. In a set of earlier studies \cite{Tero2004,hongkong,Malaysia,kim2010,Wang2003,Suh2003,Pakistan1,Pakistan2,Pakistan3,Zafar2008,Korea1},
similar investigations have been made. However, the important question:
Which technical (cryptographic) measures taken by the bank can improve
the security of e-service and thus lead to enhanced confidence among
the end users had not been discussed in the earlier works. The present
work aims to go beyond technology acceptance model and to address
this issue by highlighting the limitations of the security measures
taken by different banks.

The rest of the paper is organized as follows. In Section \ref{sec:Vulnerability-analysis},
we perform the vulnerability analysis of Indian banks and establish
that the encryption used by Indian banks are often vulnerable. In
Section \ref{sec:Comparison-of-the}, we compare the encryption techniques
adopted by the Indian banks with the international standards and analyze
them in view of the recommendations of RBI. The comparison, has revealed
that many banks are not strictly following RBI guidelines and thus
prone to cyber attacks. Subsequently, in Section \ref{sec:Measures-suggested-to},
we suggest a set of measures that can be adopted by Indian banks to
avoid the use of vulnerable encryption techniques in future. Finally,
the paper is concluded in Section \ref{sec:Conclusions}.

\section{Vulnerability analysis\label{sec:Vulnerability-analysis} }

\begin{figure}
\includegraphics[scale=0.6]{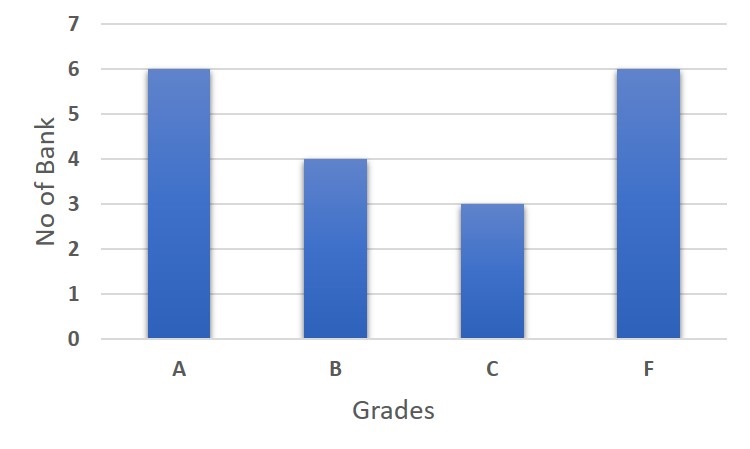}\includegraphics[scale=0.6]{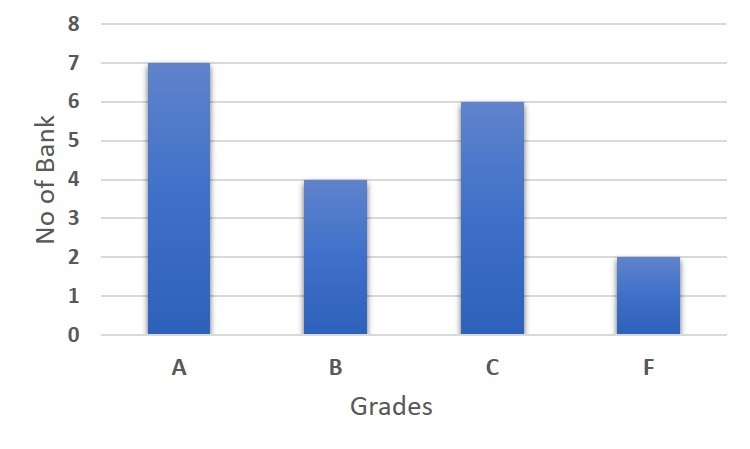}\\
 \includegraphics[scale=0.7]{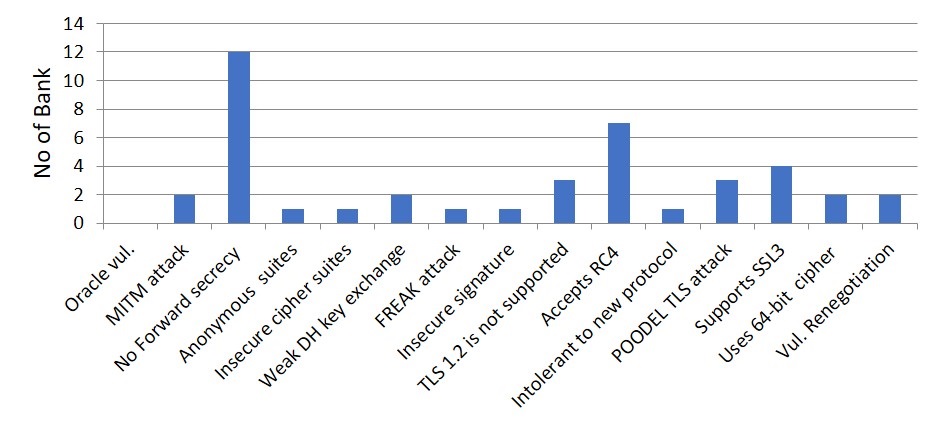}\\
\includegraphics[scale=0.6]{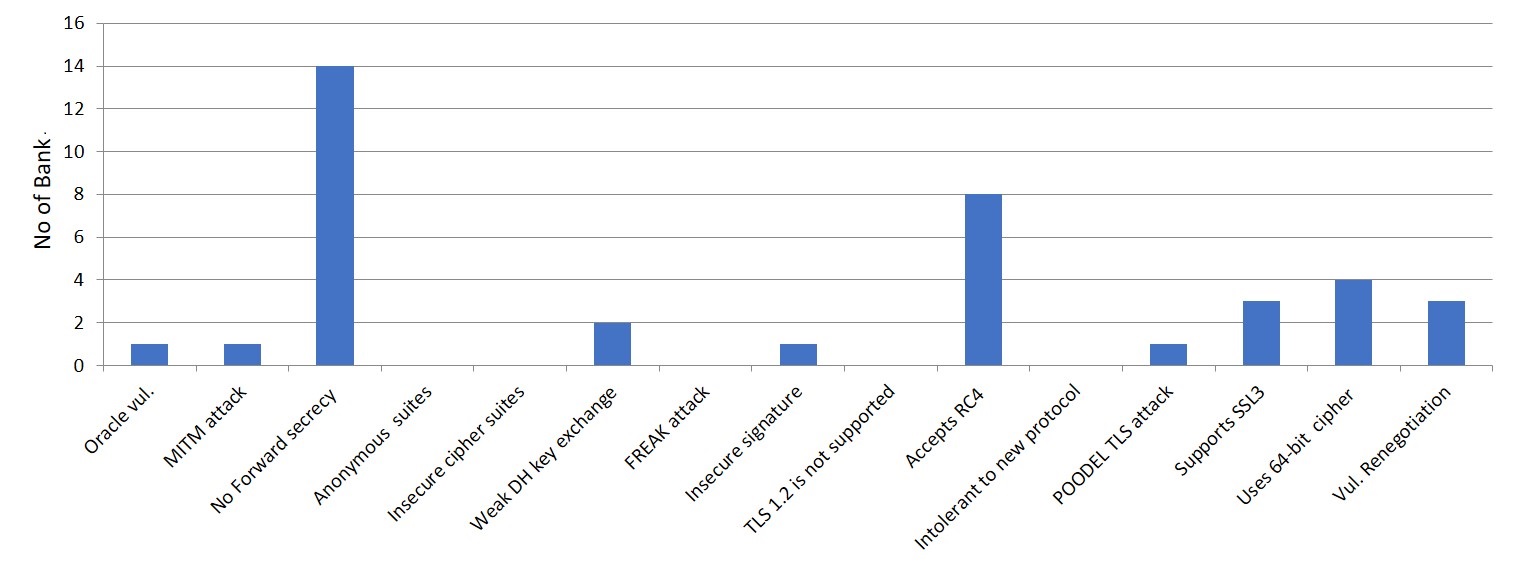}

\protect\protect\caption{\label{fig:one}(a) Number distribution of grades obtained by different
public banks, (b) Number distribution of grades obtained by different
private banks, (c) Number distribution of the Public banks showing
a particular characteristic of vulnerability (weakness), (c) Number
distribution of the Private banks showing a particular characteristic
of vulnerability (weakness).}
\end{figure}

The SSL client certificates are obtained by the banks from different
certifying agencies (CA) like, Symantec Corporation, DigiCert Inc.,
and Corporation Service Company. These organizations are allowed to
issue SSL client certificate. The online banking websites contain
details of the certificates obtained by the banks and the validity
of the certificate. It also provides certified information about the
encryption and authentication method used by the bank to encrypt information
before being transmitted over the internet. The quality of these encryption
methods can be analyzed to some extent by looking at the string like,
``TLS\_ECDHE\_RSA\_WITH\_AES\_256\_GCM\_SHA384, 256 bit keys, TLS
1.2'', which are obtained directly from the security certificate
information of the banking website. A full list of strings describing
the encryption and the authentication methods adopted by different
banks are provided in Table \ref{tab:String-that-identify} of Appendix
A. Further, information about the quality of the encryption technique
adopted and their potential vulnerability can be obtained by performing
various tests on SSL servers and/or SSL certificates. Some of the
online SSL certificate checking tools that may be used to reveal the
potential vulnerability of the adopted encryption technique are available
at \cite{ssl-labs,digi-cert,vulnarebility1,vulnarebility2}. In what
follows, we report and analyze the results of SSL server tests performed
by SSL Labs \cite{ssl-labs}. These tests not only reveal the potential
vulnerability it also grades the servers according to their strength.
A server graded as A+ seems to be most strong, whereas a server graded
as F is considered to be the weakest or most vulnerable. We have analyzed
the security of 38 banks using \cite{ssl-labs}. The result is summarized
in Fig. \ref{fig:one} (a)-(b), where \ref{fig:one} (a) provides
the number distribution of grades obtained by different Public (Government
controlled) banks and \ref{fig:one} (b) provides the same information
for the private banks. Fig. \ref{fig:one} (a)-(b), shows that 13
of the 38 banks obtained A, A+, and A- grades, and they may be viewed
to provide relatively higher security. Similarly, the analysis has
shown that B and C grades have been obtained by 8 and 9 banks respectively.
Thus, we may conclude that these banks provide moderate security.
Finally, remaining 8 banks are found to obtain F grade, and they are
highly vulnerable, implying the fact that the security measure taken
by them is very weak. There are many reasons that lead to the security
weaknesses of this banks. Such weaknesses are characterized by performing
various vulnerability tests and summarized in Tables \ref{tab:Reaonably-secure}-\ref{tab:Weally secure}.
Details of the insecure SSL ciphers supported by the server of the
banks and other drawbacks of these banks are summarized in these tables.
Specifically, Table \ref{tab:Reaonably-secure} is focused on the
analysis of security of those banks that are reasonably secure (obtains
${\rm A},$${\rm A+}$ or, ${\rm A-}$ grades). Table \ref{tab:moderate}
and Table \ref{tab:Weally secure} summarizes the security analysis
performed for the banks with moderate security (Grades B and C) and
those with weak security measures (Grade F), respectively. In a similar
line, security analysis performed for popular wallets provided by
non-Banking organizations are reported in Table \ref{tab:Security-of-wallets}.
While analyzing the limitations of the security measures taken by
a bank, in addition to the results of analysis performed by \cite{ssl-labs},
we have also used the outcome of analysis performed by \cite{digi-cert},
which essentially tests ``Heartbleed vulnerability'', checks whether
the server supports backdated protocols like TLS 1.0, TLS 1.1, SSL
3.0, etc., looks for known vulnerable Debian keys and lists all the
SSL ciphers supported by the server. This list is subsequently used
to detect whether the server supports insecure ciphers. The outputs
of analysis performed using \cite{ssl-labs,digi-cert} are combined
in Tables \ref{tab:Reaonably-secure}-\ref{tab:Security-of-wallets}.
Here it may be noted that in Table \ref{tab:Security-of-wallets},
we review the security provided by the most popular non-bank wallets
used for e-commerce by Indians. The analysis performed through \cite{ssl-labs,digi-cert}
has revealed that various banks use vulnerable ciphers, specially
some of them are found to use keys of inadequate length, support RC4
cipher with older protocols, support TLS 1.0, TLS 1.1, SSL 3.0, etc.,
and to be vulnerable to Padding Oracle On Downgraded Legacy Encryption
(POODLE) attack (Allahabad Bank, Canara Bank, Indusind Bank and Punjab
National Bank) and to man-in-the-middle attack (MITM) (Bank of Baroda,
Syndicate Bank, Indusind Bank). Here, it may be apt to note that POODLE
is a particular type of MITM attack that exploits internet and security
software clients' fallback to SSL 3.0. These observations are illustrated
through Fig. \ref{fig:one} (c), and summarized in Table \ref{tab:Bank-Issue}
of Appendix A.

%\begin{tabular}%
\begin{longtable}{|>{\centering}p{0.6cm}|>{\centering}p{1.9cm}|>{\centering}p{4.1cm}|>{\centering}p{3.9cm}|>{\raggedright}p{5cm}|}
\hline 
S.No  & Name of the bank (Pub/Pvt)  & \textbf{Website}  & \textbf{Server (Grade)}  & \textbf{Remarks} \tabularnewline
\hline 
1  & State Bank of India (Pub)  & retail.onlinesbi.com  & 223.31.160.67 \textbf{(A+)}

2405:a700:14:12c:0:0:0:148

\textbf{(A+)}  & (i) HTTP Strict Transport Security (HSTS) with long duration deployed
on this server. \tabularnewline
\hline 
2  & Axis Bank (Pvt)  & retail.axisbank.co.in  & 121.243.50.21 \textbf{(A)}

2403:0:500:11:0:0:0:160 \textbf{(A)}  & \tabularnewline
\hline 
3  & Andhra Bank (Pub)  & www.online andhrabank.net.in  & 103.196.117.1 \textbf{(A)}  & \tabularnewline
\hline 
4  & Catholic Syrian Bank (Pvt)  & www.csbnet.co.in  & 115.112.233.198 \textbf{(A)}  & \tabularnewline
\hline 
5  & HDFC Bank

(Pvt)  & netbanking.hdfcbank.com  & 175.100.160.21 \textbf{(A) }  & (i) Intermediate certificate has an insecure signature.

(ii) HTTP Strict Transport Security (HSTS) with long duration deployed
on this server. \tabularnewline
\hline 
6  & Indian Bank (Pub)  & www.indianbank.net.in  & 115.110.173.99 \textbf{(A-)}  & (i) Intermediate certificate has an insecure signature. \tabularnewline
\hline 
7  & Oriental Bank of Commerce

(Pub)  & www.obconline.co.in  & 220.226.206.37 \textbf{(A-)}  & (i) Forward Secrecy with the reference browsers is not supported.

(ii) HTTP Strict Transport Security (HSTS) with long duration deployed
on this server. \tabularnewline
\hline 
8  & UCO Bank (Pub)  & www.ucoebanking.com  & 203.200.206.227\textbf{ (A-)}

2405:a080:0:0:0:0:0:30 \textbf{(A-)}  & (i) Secure renegotiation is not supported.

(ii) Forward Secrecy with the reference browsers is not supported.\tabularnewline
\hline 
9  & United Bank of India (Pub)  & ebank.unitedbank ofindia.com  & 14.142.120.11 \textbf{(A-)}  & \tabularnewline
\hline 
10  & ICICI Bank

(Pvt)  & infinity.icicibank.com  & 203.189.92.162 \textbf{(A-)}

2001:df0:2fd:1:0:0:0:2 \textbf{(A-)}  & (i) Secure renegotiation is not supported.

(ii) Forward Secrecy with the reference browsers is not supported.\tabularnewline
\hline 
11  & Karnataka Bank (Pvt)  & moneyclick.karnataka bank.co.in  & 210.212.203.235 \textbf{(A-)}

125.16.142.105 \textbf{(A-)}

2404:a800:3001:b:0:0:0:5 \textbf{(A-)}  & (i) Forward Secrecy with the reference browsers is not supported. \tabularnewline
\hline 
12  & RBL Bank (Pvt)  & online.rblbank.com  & 180.179.110.171 \textbf{(A-)}  & (i) Secure renegotiation is not supported.

(ii) Forward Secrecy with the reference browsers is not supported.\tabularnewline
\hline 
13  & Yes Bank (Pvt)  & netbanking.yesbank.co.in  & 123.136.19.72 \textbf{(A-)}  & (i) Secure renegotiation is not supported.

(ii) Forward Secrecy with the reference browsers is not supported.\tabularnewline
\hline 
%\end{tabular}\protect
\caption{\label{tab:Reaonably-secure}List of banks that provide reasonably
good security in the sense that they obtain grades $\in\{{\rm A+,A},$~${\rm A-}\}$ }
\end{longtable}

%\begin{tabular}%
\begin{longtable}{|>{\centering}p{0.7cm}|>{\centering}p{2cm}|>{\centering}p{4cm}|>{\centering}p{4cm}|>{\raggedright}p{5cm}|}
\hline 
S. No  & Name of the bank (Pub/Pvt)  & \textbf{Website}  & \textbf{Server (Grade)}  & \textbf{Issues that reduced the grade} \tabularnewline
\hline 
1  & Vijaya Bank (Pub)  & www.vijayabankonline.in  & 210.212.204.30 \textbf{(B)}

219.65.65.166 \textbf{(B)}

2403:0:500:31:0:0:0:12 \textbf{(B)}  & (i) Accepts RC4 cipher, but only with older protocols.

(ii) Forward Secrecy with the reference browsers is not supported.\tabularnewline
\hline 
2  & Bank of India (Pub)  & starconnectcbs.bank ofindia.com  & 61.246.202.9 \textbf{(B)}  & (i) Supports weak Diffie-Hellman (DH) key exchange parameters.

(ii) Forward Secrecy with the reference browsers is not supported. \tabularnewline
\hline 
3  & Indian Overseas Bank (Pub)  & www.iobnet.co.in  & 121.242.125.68 \textbf{(B)}  & (i) Uses SSL 3, which is obsolete and insecure.

(ii) Accepts RC4 cipher, but only with older protocols.

(iii) Forward Secrecy with the reference browsers is not supported. \tabularnewline
\hline 
4  & IDBI Bank Ltd (Pub)  & inet.idbibank.co.in  & 103.93.45.42\textbf{ (B)}

2001:df1:3700:1:0:0:0:7 \textbf{(B)}  & (i) Accepts RC4 cipher, but only with older protocols.

(ii) Secure renegotiation is not supported.

(iii) Forward Secrecy with the reference browsers is not supported.\tabularnewline
\hline 
5  & DCB Bank (Pvt)  & pib.dcbbank.com  & 202.56.244.126 \textbf{(B)}

203.196.200.194 \textbf{(B)}  & (i) Accepts RC4 cipher, but only with older protocols.

(ii) Forward Secrecy with the reference browsers is not supported. \tabularnewline
\hline 
6  & Fednet Internet Banking (Pvt)  & www.fednetbank.com  & 121.243.127.68 \textbf{(B)}  & (i) Accepts RC4 cipher, but only with older protocols.

(ii) Forward Secrecy with the reference browsers is not supported. \tabularnewline
\hline 
7  & IDFC Bank (Pvt)  & my.idfcbank.com  & 103.233.77.139 \textbf{(B)}  & (i) Supports weak Diffie-Hellman (DH) key exchange parameters.

(ii) Forward Secrecy with the reference browsers is not supported. \tabularnewline
\hline 
8  & Lakshmi Vilas Bank (Pvt)  & www.lvbankonline.in  & 121.243.113.251 \textbf{(B)}  & (i) Accepts RC4 cipher, but only with older protocols.

(ii) Forward Secrecy with the reference browsers is not supported. \tabularnewline
\hline 
9  & Allahabad bank (Pub)  & www.allahabadbank.in  & 2401:8800:70:1:0:0:0:2 \textbf{(C)}

180.179.170.68 \textbf{(C)}  & (i) Vulnerable to the POODLE attack.

(ii) Supports older protocols, but does not support the current best
TLS 1.2.

(iii) Accepts RC4 cipher, but only with older protocols. \tabularnewline
\hline 
10  & Central Bank of India (Pub)  & www.centralbank.net.in  & 223.30.146.175 \textbf{(C)}

112.133.218.235 \textbf{(C)}  & (i) Vulnerable to the POODLE attack.

(ii) Accepts RC4 cipher, but only with older protocols.

(iii) Forward Secrecy with the reference browsers is not supported. \tabularnewline
\hline 
11  & Dhan Bank (Pvt)  & netbank.dhanbank.in  & 115.117.58.155 \textbf{(C)}

59.144.54.55 \textbf{(C)}  & (i) Uses SSL 3, which is obsolete and insecure.

(ii) Uses RC4 with modern protocols.

(iii) Forward Secrecy with the reference browsers is not supported. \tabularnewline
\hline 
12  & Punjab And Sind Bank

(Pub)  & www.psbonline.co.in  & 202.191.179.43 \textbf{(C)}  & (i) Accepts RC4 cipher, but only with older protocols.

(ii) Uses 64-bit block cipher (3DES / DES / RC2 / IDEA) with modern
protocols.

(iii) Forward Secrecy with the reference browsers is not supported. \tabularnewline
\hline 
13  & J \& K Bank (Pvt)  & www.jkbankonline.com  & 223.30.216.170 \textbf{(C)}  & (i) Supports weak DH key exchange parameters.

(ii) Uses SSL 3, which is obsolete and insecure.

(iii) Uses RC4 with modern protocols.

(iv) Forward Secrecy with the reference browsers is not supported. \tabularnewline
\hline 
14  & Karur Vysya Bank (Pvt)  & www.kvbnet.co.in  & 115.249.239.101 \textbf{(C)}  & (i) Accepts RC4 cipher, but only with older protocols.

(ii) Uses 64-bit block cipher (3DES / DES / RC2 / IDEA) with modern
protocols.

(iii) Forward Secrecy with the reference browsers is not supported. \tabularnewline
\hline 
15  & Kotak Mahindra Bank (Pvt)  & www.kotak.com  & 203.196.200.28 \textbf{(C)}

2403:0:100:51:0:0:0:51 \textbf{(C)}  & (i) Uses RC4 with modern protocols.

(ii) Uses 64-bit block cipher (3DES / DES / RC2 / IDEA)

(iii) Forward Secrecy with the reference browsers is not supported. \tabularnewline
\hline 
16  & South Indian Bank (Pvt)  & sibernet.southindian bank.com  & 103.212.29.50 \textbf{(C)}  & (i) Uses 64-bit block cipher (3DES / DES / RC2 / IDEA) with modern
protocols.

(ii) Forward Secrecy with the reference browsers is not supported. \tabularnewline
\hline 
17  & Tamilnad Mercantile Bank (Pvt)  & www.tmbnet.in  & 124.124.113.139 \textbf{(A)} 14.141.76.115 \textbf{(A)} 115.254.82.230
\textbf{(C)} 14.142.204.70 \textbf{(C)}  & \tabularnewline
\hline 
%\end{tabular}%\protect
\caption{\label{tab:moderate}List of banks that provide moderate security
in the sense that they obtain grades $\in\{{\rm B,C}$\}. }

\end{longtable}

%\begin{tabular}%
\begin{longtable}{|>{\centering}p{0.7cm}|>{\centering}p{2cm}|>{\centering}p{4cm}|>{\centering}p{4cm}|>{\raggedright}p{5cm}|}
\hline 
S. No  & Name of the bank (Pub/Pvt)  & \textbf{Website}  & \textbf{Server(s) (Grade)}  & \textbf{Issues that reduced the grade} \tabularnewline
\hline 
1  & Canara Bank (Pub)  & netbanking.canarabank.in  & 180.92.164.8 \textbf{(F)}

2404:a500:0:1000:0:0:60:e\textbf{ (F)}  & (i) Vulnerable to MITM attacks because it supports insecure renegotiation.

(ii) Forward Secrecy with the reference browsers is not supported. \tabularnewline
\hline 
2  & Corporation Bank (Pub)  & www.corpretail.com  & 202.62.247.23 \textbf{(F)}  & (i) Supports anonymous (insecure) suites and other insecure cipher
suites

(ii) Supports weak DH key exchange parameters. \tabularnewline
\hline 
3  & Dena Bank (Pub)  & www.denaiconnect.co.in  & 103.224.110.30 \textbf{(F)}  & (i) Supports insecure cipher suites (see below for details).

(ii) Supports 512-bit export suites and might be vulnerable to the
FREAK attack.

(iii) Intermediate certificate contains an insecure signature. (iv)
Supports older protocols, but does not support the current best TLS
1.2.

(v) Forward Secrecy with the reference browsers is not supported.

(vi) Accepts RC4 cipher, but only with older protocols. \tabularnewline
\hline 
4  & Punjab National Bank (Pub)  & netbanking.netpnb.com  & 103.59.140.7 \textbf{(F)}

223.31.51.214 \textbf{(F)}  & (i) Vulnerable to the POODLE TLS attack.

(ii) Intolerant to newer protocol versions, which might cause connection
failures.

(iii) Supports older protocols, but does not support the current best
TLS 1.2.

(iv) Forward Secrecy with the reference browsers is not supported.\tabularnewline
\hline 
5  & Syndicate Bank (Pub)  & www.syndonline.in  & 220.226.205.223 \textbf{(F)}  & (i) Vulnerable to MITM attacks because it supports insecure renegotiation.

(ii) Uses SSL 3, which is obsolete and insecure.

(iii) Uses RC4 with modern protocols.

(iv) Uses 64-bit block cipher (3DES / DES / RC2 / IDEA) with modern
protocols.

(v) Forward Secrecy with the reference browsers is not supported. \tabularnewline
\hline 
6  & Indusind (Pvt)  & indusnet.indusind.com  & 203.196.200.211 \textbf{(F)}  & (i) Vulnerable to the POODLE attack.

(ii) Vulnerable to MITM attacks because it supports insecure renegotiation.

(iii) Forward Secrecy with the reference browsers is not supported. \tabularnewline
\hline 
7  & Bhandhan Bank (Pvt)  & bandhanbankonline.com  & 103.231.79.57 \textbf{(F)}

2400:3b00:20:4:0:0:0:92 \textbf{(F)}  & (i) Vulnerable to the OpenSSL Padding Oracle vulnerability (CVE-2016-2107)
and thus insecure.

(ii) HTTP Strict Transport Security (HSTS) with long duration deployed
on this server. \tabularnewline
\hline 
8  & Bank of Baroda (Pub)  & www.bobibanking.com  & 14.140.233.71 \textbf{(F)}

2001:e48:22:100a:0:0:0:11  & \tabularnewline
\hline 
%\end{tabular}\protect
\caption{\label{tab:Weally secure}List of banks that provide reasonably weak
security in the sense that they have obtained F grades. }
\end{longtable}

%\begin{tabular}%
\begin{longtable}{|>{\centering}p{0.7cm}|>{\centering}p{2cm}|>{\centering}p{3cm}|>{\centering}p{5cm}|>{\centering}p{5cm}|}
\hline 
S. No  & Name of the bank  & \textbf{Website}  & \textbf{Server (Grade)}  & \textbf{Issues that reduced the grade} \tabularnewline
\hline 
1  & Paytm payments bank  & paytm.com  & 52.221.159.215 \textbf{(A+)} 52.77.12.221 \textbf{(A+)}  & HTTP Strict Transport Security (HSTS) with long duration deployed
on this server. \tabularnewline
\hline 
2  & Airtel payments bank  & www.airtel.in  & 23.13.173.139 \textbf{(B)}  & This server accepts RC4 cipher, but only with older protocols. Grade
capped to B. \tabularnewline
\hline 
3  & Freecharge  & www.freecharge.in  & 2600:1408:10:1a8:0:0:0:1c64 \textbf{(A)}

23.49.179.215 \textbf{(A)} 2600:1408:10:1ae:0:0:0:1c64 \textbf{(A)}  & \tabularnewline
\hline 
4  & Mobikwik  & www.mobikwik.com  & 180.179.23.136 \textbf{(A)} 2401:8800:c11:3:0:0:0:4 \textbf{(A)}  & \tabularnewline
\hline 
%\end{tabular}\protect
\caption{\label{tab:Security-of-wallets}Security analysis of the wallets that
are popularly used in India.}
\end{longtable}

\section{Comparison of the encryption techniques used with the benchmark adopted
internationally and advised by the regulatory authority \label{sec:Comparison-of-the}}

In Point 6 of RBI circular \cite{RBI-1}, it is categorically mentioned
in the context of arrangement for continuous surveillance, that ``Testing
for vulnerabilities at reasonable intervals of time is very important.
The nature of cyber-attacks are such that they can occur at any time
and in a manner that may not have been anticipated. Hence, it is mandated
that a SOC (Security Operations Centre) be set up at the earliest,
if not yet been done. It is also essential that this Centre ensures
continuous surveillance and keeps itself regularly updated on the
latest nature of emerging cyber threats''. The analysis performed
above clearly indicates that most of the Indian banks are not following
this recommendation of RBI as they are still using vulnerable and
backdated ciphers. Further, in the same context (i.e., in the context
of vulnerability), in Ref. \cite{RBI-2} Page 31, Section 16, RBI
has clearly emphasized on the necessity of regular vulnerability tests
by stating the following, ``... Banks that do not scan for vulnerabilities
and address discovered flaws proactively face a significant likelihood
of having their computer systems compromised. ii. The following are
some of the measures suggested:

``a. Automated vulnerability scanning tools need to be used against
all systems on their networks on a periodic basis, say monthly or
weekly or more frequently.

``b. Banks should ensure that vulnerability scanning is performed
in an authenticated mode (i.e., configuring the scanner with administrator
credentials) at least quarterly, ...''.

It's clear that many banks are not strictly following this advise
of RBI. In Ref. \cite{RBI-2} Page 29, Section 14, point (v), in the
context of encryption, RBI has advised banks as follows, ``Normally,
a minimum of 128-bit SSL encryption is expected. Constant advances
in computer hardware, cryptanalysis and distributed brute force techniques
may induce use of larger key lengths periodically. It is expected
that banks will properly evaluate security requirements associated
with their internet banking systems and other relevant systems and
adopt an encryption solution that is commensurate with the degree
of confidentiality and integrity required. Banks should only select
encryption algorithms which are well established international standards
and which have been subjected to rigorous scrutiny by an international
cryptographer community....''. Now although RBI is recommending a
minimum of 128-bit SSL encryption, it is found that Kayur Vashiya
Bank, Dena Bank\footnote{Recently stopped using 112 bit encryption},
Kotak Mahinrda, The South Indian Bank, Punjab and Sind bank, etc.,
are still providing 112 bit encryption which is bellow the recommended
norm. If we further concentrate on the last part of the RBI advise
that recommends the use of ``encryption algorithms which are well
established international standards and which have been subjected
to rigorous scrutiny by an international cryptographer community'',
we would easily realize that Bank should now follow at least 256-bit
encryption to be consistent with the international standards (i.e., recommendations of NIST,
ENISA and IETF\footnote{IETF is the body that is authorized to approve internet standards
and protocols. The recommendation of IETF for the algorithms to be
used for practical application are valid only for a period of six
months, and consequently, it's important to update the security measures
taken by a bank at least once in every six months.} \cite{NIST2-2017, NIST1-2011, forENISA1}) and stop supporting SHA, SHA-1, SHA-224,
TLS 1.0. TLS 1.1 and other vulnerable ciphers mentioned above. In
fact banks should now start using TLS 1.3 as it is faster than TLS
1.2 and it has many advantages over TLS 1.2. Further, recently IETF
has approved TLS 1.3 as the next version of TLS protocols \cite{TLS13}.

\section{Measures suggested to circumvent the vulnerability of encryption
techniques in future \label{sec:Measures-suggested-to} }

Banks should immediately stop supporting vulnerable ciphers. In the
next phase, they should implement post-quantum crytographic protocols
\cite{post-qua} like, lattice-based, multivariate, code-based, hash-based
cryptography, and replace the random number generators by quantum
random number generators \cite{qrng} and thus go for a hybrid (classical-quantum)
technology. Subsequently, they should look at the possibilities of
implementing semi-quantum \cite{SQ,QPC} protocols which would require
end users (customers) to have quantum resources which are costly.
A final goal should be to implement unconditionally secure and device
independent quantum schemes \cite{QC,lo}.

\section{Conclusions\label{sec:Conclusions}}

Most of the Indian banks are found to support insecure SSL ciphers.
Some of the previous studies were restricted to the banks of specific
region or country (e.g., investigations performed in Refs \cite{Tejinder,hongkong,Malaysia,Pakistan1,Pakistan2,Pakistan3,Korea1}
were restricted to the banks of India, Hongkong and Malyasia, respectively),
whereas others were not restricted. Present study is also focused
on Indian banks, but the methodology adopted and the conclusions obtained
are valid in general. To emphasis on this point we analyze the security
of 11 non-Indian banks, which are selected from different continents.
In what follows, we briefly note the outcome of such an investigation
in Table \ref{tab:Bank-Issue;-add}.

%\begin{tabular}%
\begin{longtable}{|>{\centering}p{0.7cm}|>{\centering}p{1.8cm}|>{\centering}p{3cm}|>{\centering}p{3.7cm}|>{\centering}p{6cm}|}
\hline 
\textbf{S. No}  & Name of the bank (Country)  & \textbf{Website}  & \textbf{Server (Grade)}  & \textbf{Issues that reduced the grade} \tabularnewline
\hline 
1  & Bank of America (USA)  & https://secure.bank ofamerica.com/  & 171.161.203.200 \textbf{(A-)}  & This server does not support Forward Secrecy with the reference browsers.
Grade will be capped to B from March 2018. This server's certificate
will be distrusted by Google and Mozilla from September 2018. HTTP
Strict Transport Security (HSTS) with long duration deployed on this
server.\tabularnewline
\hline 
2  & jpmorgan chase (USA)  & www.chase.com  & 159.53.224.21 \textbf{(A-)}  & This server does not support Forward Secrecy with the reference browsers.
Grade will be capped to B from March 2018. HTTP Strict Transport Security
(HSTS) with long duration deployed on this server.\tabularnewline
\hline 
3  & Bank of Brazil (Brazil)  & www.bb americas.com  & 52.84.237.182 \textbf{(A)} 52.84.237.72 \textbf{(A)} 52.84.237.12
\textbf{(A)} 52.84.237.180 \textbf{(A)} 52.84.237.121 \textbf{(A)}
52.84.237.77 \textbf{(A)} 52.84.237.80 \textbf{(A)} 52.84.237.203
\textbf{(A)}  & \tabularnewline
\hline 
4  & Steward Bank (Zimbabwe)  & onlinebanking. stewardbank.co.zw  & 41.216.125.233 \textbf{(B)}  & This server supports weak Diffie-Hellman (DH) key exchange parameters.
Grade capped to B. This site works only in browsers with SNI support. \tabularnewline
\hline 
5  & Avtovaz bank (Russia)  & server29.cey-ebanking.com  & 12.191.20.25 \textbf{(C)}  & The server supports only older protocols, but not the current best
TLS 1.2. Grade capped to C. This server does not support Forward Secrecy
with the reference browsers. Grade will be capped to B from March
2018. This server does not support Authenticated encryption (AEAD)
cipher suites. Grade will be capped to B from March 2018.\tabularnewline
\hline 
6  & Pingan Bank (China)  & bank.pingan.com  & 115.231.227.16\textbf{ (A)} 27.148.164.23\textbf{(A)}  & \tabularnewline
\hline 
7  & Police Bank Ltd (Australia)  & ebanking.police bank.com.au  & 103.11.142.65 \textbf{(A)}  & This server's certificate will be distrusted by Google and Mozilla
from September 2018\tabularnewline
\hline 
8  & Bank of Melbourne (Australia)  & ibanking.bankof melbourne.com.au  & 203.23.44.204 \textbf{(A)}  & This server does not support Authenticated encryption (AEAD) cipher
suites. Grade will be capped to B from March 2018.\tabularnewline
\hline 
9  & Bank Kerjasama Rakyat Malaysia Berhad (Malaysia)  & www2.irakyat .com.my  & 1.9.61.247 \textbf{(B)}  & This server is vulnerable to the Return of Bleichenbacher's Oracle
Threat (ROBOT) vulnerability. Grade will be set to F from March 2018.
This server supports weak Diffie-Hellman (DH) key exchange parameters.
Grade capped to B. This server uses SSL 3, which is obsolete and insecure.
Grade capped to B. This server accepts RC4 cipher, but only with older
protocols. Grade capped to B. This server does not support Forward
Secrecy with the reference browsers. Grade will be capped to B from
March 2018\tabularnewline
\hline 
10  & National Bank of Pakistan

(Pakistan)  & https://nbp.com. pk/login/  & 167.114.191.212 \textbf{(A+)}  & This server's certificate will be distrusted by Google and Mozilla
from September 2018. HTTP Strict Transport Security (HSTS) with long
duration deployed on this server.\tabularnewline
\hline 
11  & Agrani Bank Limited (Bangladesh)  & www.agranibank.org  & 180.92.224.101 \textbf{(B)}  & This server supports weak Diffie-Hellman (DH) key exchange parameters.
Grade capped to B. This server does not support Forward Secrecy with
the reference browsers. Grade will be capped to B from March 2018.
This server's certificate chain is incomplete. Grade capped to B.
This server's certificate will be distrusted by Google and Mozilla
from September 2018. \tabularnewline
\hline 
%\end{tabular}\protect
\caption{\label{tab:Bank-Issue;-add}Security analysis of some non-Indian banks. }
\end{longtable}

This study is unique, as it's the first one of its kind in the context
of Indian banks. However, the analysis performed is not deep enough.
Consequently, it opens up the possibility of a deeper analysis in
future (which will be reported elsewhere). As the adopted method is
valid in general, it's possible to use this simple approach to analyze
the security of banks and other e-commerce sites of other countries,
too. Finally, we conclude the article with an optimistic view and
a hope that this article will be able to draw the attention of the
concerned authorities, and necessary steps will be taken by them to
perform regular checking of vulnerability of the encryption techniques
adopted by then (as recommended by RBI) and to take corrective measures
wherever required. This would help banks to provide higher security
and thus to improve the confidence level of the end-users (acceptability
of the end-users), which the banks desire to improve. Further, we
hope that the suggestions made from the perspectives of a cryptographer
will be used by the banks in near future and that would help them
to circumvent many attacks in the future.

\textbf{Acknowledgment:} AP thanks Defense Research and Development
Organization (DRDO), India for the support provided through the Project
Number: ERIP/ER/1403163/M/01/1603. Authors also thank Abhishek Parakh and Kishore Thapliyal for their
interest in this work and some useful technical comments.

\pagebreak

\section*{Appendix A}
%dummy comment inserted by tex2lyx to ensure that this paragraph is not empty
\setcounter{table}{0}
\global\long\def\thetable{A.\Roman{table}}

%\begin{tabular}%
\begin{longtable}{|>{\centering}p{0.5cm}|>{\centering}p{2cm}|>{\centering}p{3.5cm}|>{\centering}p{7.5cm}|>{\centering}p{2cm}|}
\hline 
\textbf{S. No}  & \textbf{Bank Name}  & \textbf{Website}  & \textbf{Connection Encrypted}  & \textbf{Verified by (Certification Authority)}\tabularnewline
\hline 
1  & Allahabad Bank  & www.allahabadbank.in  & TLS\_ECDHE\_RSA\_WITH\_AES\_256\_CBC\_ SHA, 256 bit keys, TLS 1.0  & Symantec Corporation\tabularnewline
\hline 
2  & Andhra Bank  & www.onlineandhra bank.net.in  & TLS\_ECDHE\_RSA\_WITH\_AES\_256\_GCM\_ SHA384, 256 bit keys, TLS 1.2  & Entrust, Inc.\tabularnewline
\hline 
3  & State Bank of India  & retail.onlinesbi.com  & TLS\_ECDHE\_RSA\_WITH\_AES\_256\_GCM\_ SHA384, 256 bit keys, TLS 1.2  & Symantec Corporation\tabularnewline
\hline 
4  & Axis Bank Limited  & retail.axisbank.co.in  & TLS\_ECDHE\_RSA\_WITH\_AES\_256\_GCM\_ SHA384, 256 bit keys, TLS 1.2  & Symantec Corporation\tabularnewline
\hline 
5  & HDFC Bank Limited.  & netbanking. hdfcbank.com  & TLS\_ECDHE\_RSA\_WITH\_AES\_256\_GCM\_ SHA384, 256 bit keys, TLS 1.2  & Symantec Corporation\tabularnewline
\hline 
6  & Jammu and Kashmir Bank Limited  & starconnectcbs. bankofindia.com  & TLS\_ECDHE\_RSA\_WITH\_AES\_256\_GCM\_ SHA384, 256 bit keys, TLS 1.2  & Symantec Corporation\tabularnewline
\hline 
7  & Bandhan Bank Limited  & bandhanbankonline.com  & TLS\_ECDHE\_RSA\_WITH\_AES\_256\_GCM\_ SHA384, 256 bit keys, TLS 1.2  & Entrust, Inc.\tabularnewline
\hline 
8  & Vijaya Bank  & www.vijayabank online.in  & TLS\_ECDHE\_RSA\_WITH\_AES\_256\_CBC\_ SHA, 256 bit keys, TLS 1.2  & Entrust, Inc.\tabularnewline
\hline 
9  & The Catholic Syrian Bank Ltd  & www.csbnet.co.in  & TLS\_ECDHE\_RSA\_WITH\_AES\_256\_CBC\_ SHA, 256 bit keys, TLS 1.2  & Symantec Corporation\tabularnewline
\hline 
10  & Tamilnad Mercantile Bank Limited  & www.tmbnet.in  & TLS\_ECDHE\_RSA\_WITH\_AES\_256\_CBC\_ SHA, 256 bit keys, TLS 1.2  & Symantec Corporation\tabularnewline
\hline 
11  & Canara Bank  & netbanking.canara bank.in  & TLS\_RSA\_WITH\_AES\_256\_CBC\_ SHA, 256 bit keys, TLS 1.2  & GlobalSign nv-sa\tabularnewline
\hline 
12  & Corporation Bank  & www.corpretail.com  & TLS\_RSA\_WITH\_AES\_256\_CBC\_ SHA, 256 bit keys, TLS 1.2  & GlobalSign nv-sa\tabularnewline
\hline 
13  & Indian Overseas Bank  & www.iobnet.co.in  & TLS\_RSA\_WITH\_AES\_256\_CBC\_ SHA, 256 bit keys, TLS 1.2  & Symantec Corporation\tabularnewline
\hline 
14  & Oriental Bank of Commerce  & www.obconline.co.in  & TLS\_RSA\_WITH\_AES\_256\_CBC\_ SHA, 256 bit keys, TLS 1.2  & Symantec Corporation\tabularnewline
\hline 
15  & IDBI Bank Ltd  & inet.idbibank.co.in  & TLS\_RSA\_WITH\_AES\_256\_CBC\_ SHA, 256 bit keys, TLS 1.2  & Entrust, Inc.\tabularnewline
\hline 
16  & United Bank of India & ebank.unitedbank ofindia.com  & TLS\_RSA\_WITH\_AES\_256\_CBC\_ SHA, 256 bit keys, TLS 1.2  & Symantec Corporation\tabularnewline
\hline 
17  & Dhanlaxmi Bank Limited  & netbank.dhanbank.in  & TLS\_RSA\_WITH\_AES\_256\_CBC\_ SHA, 256 bit keys, TLS 1.2  & Entrust, Inc.\tabularnewline
\hline 
18  & ICICI Bank Limited  & infinity.icicibank.com  & TLS\_RSA\_WITH\_AES\_256\_CBC\_ SHA, 256 bit keys, TLS 1.2  & Entrust, Inc.\tabularnewline
\hline 
19  & IndusInd Bank Ltd  & indusnet.indusind.com  & TLS\_RSA\_WITH\_AES\_256\_CBC\_ SHA, 256 bit keys, TLS 1.2  & Entrust, Inc.\tabularnewline
\hline 
20  & RBL Bank Limited  & online.rblbank.com  & TLS\_RSA\_WITH\_AES\_256\_CBC\_ SHA, 256 bit keys, TLS 1.2  & Entrust, Inc.\tabularnewline
\hline 
21  & Yes Bank Limited  & netbanking.yes bank.co.in  & TLS\_RSA\_WITH\_AES\_256\_CBC\_ SHA, 256 bit keys, TLS 1.2  & Symantec Corporation\tabularnewline
\hline 
22  & PUNJAB AND SIND BANK  & www.psbonline.co.in  & TLS\_RSA\_WITH\_3DES\_EDE\_CBC\_ SHA, 112 bit keys, TLS 1.2  & GlobalSign nv-sa\tabularnewline
\hline 
23  & Syndicate Bank  & www.syndonline.in  & TLS\_RSA\_WITH\_3DES\_EDE\_CBC\_ SHA, 112 bit keys, TLS 1.2  & Symantec Corporation\tabularnewline
\hline 
24  & IDFC Bank Limited  & my.idfcbank.com  & TLS\_RSA\_WITH\_3DES\_EDE\_CBC\_ SHA, 112 bit keys, TLS 1.2  & Entrust, Inc.\tabularnewline
\hline 
25  & Kotak Mahindra Bank Ltd  & www.kotak.com  & TLS\_RSA\_WITH\_3DES\_EDE\_CBC\_ SHA, 112 bit keys, TLS 1.2  & Entrust, Inc.\tabularnewline
\hline 
26  & The Karur Vysya Bank Ltd  & www.kvbnet.co.in  & TLS\_RSA\_WITH\_3DES\_EDE\_CBC\_ SHA, 112 bit keys, TLS 1.2  & Symantec Corporation\tabularnewline
\hline 
27  & The South Indian Bank Ltd  & sibernet.southindian bank.com  & TLS\_RSA\_WITH\_3DES\_EDE\_CBC\_ SHA, 112 bit keys, TLS 1.2  & Entrust, Inc\tabularnewline
\hline 
28  & Bank of Baroda  & www.bobibanking.com  & TLS\_RSA\_WITH\_AES\_128\_CBC\_ SHA, 128 bit keys, TLS 1.0  & Symantec Corporation\tabularnewline
\hline 
29  & Punjab National Bank  & netbanking.netpnb.com  & TLS\_RSA\_WITH\_AES\_128\_CBC\_ SHA, 128 bit keys, TLS 1.0  & Symantec Corporation\tabularnewline
\hline 
30  & Indian Bank  & www.indianbank.net.in  & TLS\_RSA\_WITH\_AES\_128\_CBC\_ SHA, 128 bit keys, TLS 1.2  & Symantec Corporation\tabularnewline
\hline 
31  & UCO Bank  & www.ucoebanking.com  & TLS\_RSA\_WITH\_AES\_128\_CBC\_ SHA, 128 bit keys, TLS 1.2  & Symantec Corporation\tabularnewline
\hline 
32  & The Federal Bank Limited  & www.fednetbank.com  & TLS\_RSA\_WITH\_AES\_128\_CBC\_ SHA, 128 bit keys, TLS 1.2  & Symantec Corporation\tabularnewline
\hline 
33  & Karnataka Bank Limited  & moneyclick.karnataka bank.co.in  & TLS\_RSA\_WITH\_AES\_128\_CBC\_ SHA, 128 bit keys, TLS 1.2  & Symantec Corporation\tabularnewline
\hline 
34  & LAKSHMI VILAS BANK LIMITED  & www.lvbankonline.in  & TLS\_RSA\_WITH\_AES\_128\_CBC\_ SHA, 128 bit keys, TLS 1.2  & Symantec Corporation\tabularnewline
\hline 
35  & DCB BANK  & pib.dcbbank.com  & TLS\_RSA\_WITH\_AES\_128\_CBC\_S HA, 128 bit keys, TLS 1.2  & Symantec Corporation\tabularnewline
\hline 
36  & Bank of India  & starconnectcbs.bank ofindia.com  & TLS\_DHE\_RSA\_WITH\_AES\_256\_CBC\_ SHA, 256 bit keys, TLS 1.2  & Entrust, Inc.\tabularnewline
\hline 
37  & Central Bank of India  & www.centralbank.net.in  & TLS\_ECDHE\_RSA\_WITH\_AES\_128\_CBC\_ SHA, 128 bit keys, TLS 1.2  & Symantec Corporation\tabularnewline
\hline 
38  & Dena Bank  & www.denaiconnect.co.in  & TLS\_RSA\_WITH\_3DES\_EDE\_CBC\_ SHA, 112 bit keys, TLS 1.0  & Symantec Corporation\tabularnewline
\hline 
%\end{tabular}\protect
\caption{\label{tab:String-that-identify}String that identifies the security
measures taken by a bank for encryption and authentication is listed
along with the name of CA who has issued the security certificate.}
\end{longtable}

%\begin{tabular}%
\begin{longtable}{|>{\centering}p{0.5cm}|>{\centering}p{5cm}|>{\centering}p{2cm}|>{\centering}p{2cm}|>{\centering}p{2cm}|}
\hline 
S. No  & Weakness  & Public Bank  & Private Bank  & Total\tabularnewline
\hline 
1  & vulnerable to the OpenSSL Padding Oracle vulnerability  & 0  & 1  & 1\tabularnewline
\hline 
2  & MITM attack  & 2  & 1  & 3\tabularnewline
\hline 
3  & Forward secrecy with the reference browsers is not suported  & 12  & 14  & 26\tabularnewline
\hline 
4  & supports anonymous (insecure) suites  & 1  & 0  & 1\tabularnewline
\hline 
5  & supports insecure cipher suites  & 1  & 0  & 1\tabularnewline
\hline 
6  & supports weak DH key exchange  & 2  & 2  & 4\tabularnewline
\hline 
7  & FREAK attack  & 1  & 0  & 1\tabularnewline
\hline 
8  & intermediate ceritificate with insecure signature  & 1  & 1  & 2\tabularnewline
\hline 
9  & TLS 1.2 is not supported  & 3  & 0  & 3\tabularnewline
\hline 
10  & Accepts/uses RC4  & 7  & 8  & 15\tabularnewline
\hline 
11  & Intolerant to newer protocol  & 1  & 0  & 1\tabularnewline
\hline 
12  & POODEL TLS attack  & 3  & 1  & 4\tabularnewline
\hline 
13  & Supports SSL3  & 4  & 3  & 7\tabularnewline
\hline 
14  & uses 64-bit block cipher (3DES / DES / RC2 / IDEA) with modern protocols  & 2  & 4  & 6\tabularnewline
\hline 
15  & No support for secure renegotiation.  & 2  & 3  & 5\tabularnewline
\hline 
%\end{tabular}\protect
\caption{\label{tab:Bank-Issue} Various vulnerable ciphers that are used by
the bank with low grades.}
\end{longtable}

\begin{thebibliography}{10}
\bibitem{RBI-1}Cyber Security Framework in Banks (2016), Issued by RBI through
circular number RBI/2015-16/418; DBS.CO/CSITE/BC.11/33.01.001/2015-16
and available at https://www.rbi.org.in/scripts/BS\_CircularIndexDisplay.aspx?Id=10435

\bibitem{RBI-2}Guidelines on Information security, Electronic Banking,
Technology risk management and cyber frauds. Issued by RBI and available
at https://rbidocs.rbi.org.in/rdocs/content/PDFs/GBS300411F.pdf

\bibitem{moondra}S. S. Mundra, Information technology and cyber risk in banking sector-
the emerging fault lines, International Seminar on Cyber Risk and Mitigation for banks, organized by the Centre for 
Advanced Financial Research and Learning (CAFRAL), Mumbai, 7 September 2016. https://www.bis.org/review/r160909a.pdf

\bibitem{Tejinder}T. Singh,  Security and
Privacy Issues in E-Banking: An Empirical Study of Customers' Perception.
A project report submitted to Indian Institute of Banking and Finance
(2013) and available at http://www.iibf.org.in/documents/reseach-report/Tejinder\_Final\%20.pdf

\bibitem{Bhutt-2011-India}S. C. Bhutt, Study of Indian
Banks Websites for Cyber Crime Safety Mechanism.
International Journal of Advanced Computer Science and Applications,
\textbf{2} (2011) 87. Retrieved from http://thesai.org/Downloads/Volume2No10/Paper\%2014-Stu
dy\%20of\%20Indian\%20Banks\%20Websites\%20for\%20Cyb er\%20Crime\%20Safety\%20Mechanism.pdf

\bibitem{Gupta07} P. K. Gupta,  Internet banking
in India\textendash Consumer concerns and bank strategies.
Proceedings of Global Conference on Business and Finance, May 23-26,
2007, San Jose, Costa Rica,\textbf{2} (2007) 59. Also in Global Journal of Business Research \textbf{2} (2008) 43.

\bibitem{Hamed2017} H. Taherdoost, Understanding
of e-service security dimensions and its effect on quality and intention
to use. Information and Computer Security.  \textbf{25}
(2017) 535.

\bibitem{Hameed2018} H. Taherdoost, Development
of an adoption model to assess user acceptance of e-service technology:
E-Service Technology Acceptance Model. Behaviour and
Information Technology.  \textbf{37}  (2018) 173.

\bibitem{Tero2004}  T. Pikkarainen, K. Pikkarainen, H. Karjaluoto and  S.Pahnila,  Consumer acceptance of online banking:
an extension of the technology acceptance model,  Internet
Research. \textbf{14} (2004) 224.

\bibitem{hongkong} T. E. Cheng,  D. Y. Lam and A. C. Yeung,
Adoption of internet banking: an empirical study in
Hong Kong, Decision support systems. \textbf{42} (2006) 1558.

\bibitem{Malaysia} M. Lallmahamood, An Examination
of Individual's Perceived Security and Privacy of the Internet in
Malaysia and the Influence of this on their Intention to Use E-commerce:
Using an Extension of the Technology Acceptance Model,
Journal of Internet Banking and Commerce. \textbf{12} (2007) 1.

\bibitem{kim2010}  C. Kim, W. Tao, N. Shin and K. S. Kim, An empirical study of customers' perceptions of
security and trust in e-payment systems, Electronic
commerce research and applications. \textbf{9} (2010) 84.

\bibitem{Wang2003} Y. S. Wang, Y. M. Wang, H. H.Lin and T. I. Tang, Determinants of user acceptance of Internet banking:
an empirical study, International journal of service
industry management.  \textbf{14} (2003) 501.

\bibitem{Suh2003} B. Suh and I. Han, The impact
of customer trust and perception of security control on the acceptance
of electronic commerce, International Journal of electronic
commerce. \textbf{7} (2003) 135.

\bibitem{Pakistan1}  A. Manzoor, Protecting Customers Online: Response
from Pakistani Banks, International Journal of Science and Applied
Information Technology. \textbf{3} January - February issue (2014) 1; Available
Online at http://warse.org/pdfs/2014/ijsait01312014.pdf

\bibitem{Pakistan2}S. A. Raza and N. Hanif, Factors
affecting internet banking adoption among internal and external customers:
a case of Pakistan. International Journal of Electronic
Finance \textbf{7} (2013) 82.

\bibitem{Pakistan3}A. Omar,  N. Sultan, K. Zaman, N.
Bibi, A. Wajid and K. Khan, Customer perception
towards online banking services: Empirical evidence from Pakistan \textbf{16} (2011).

\bibitem{Zafar2008} T. M. Qureshi,  K. Zafar and B. Khan, Customer acceptance
of online banking in developing economies, Journal
of Internet Banking and Commerce \textbf{13} (2008) 1.

\bibitem{Korea1}H. Kim,  J. H. Huh and R. Anderson, On the security of internet banking in South Korea, (2010). Available Online at https://ora.ox.ac.uk/catalog/uuid:e3cf724a-ab9a-4f5a-87d0-b028e58fac7a/download\_file?file\_format=pdf\&safe\_filename=RR-10-01.pdf\&type\_of\_work=Report.


\bibitem{ssl-labs}https://www.ssllabs.com/ssltest/ used on 9-10 October
2017 and on February 23, 2018; Grades and issues reported here are
found on February 23, 2018. This is a free online service that claimed
to perform a deep analysis of the configuration of any SSL web server
on the public Internet.

\bibitem{digi-cert}https://www.digicert.com/help/ used on used on
9-10 October 2017 and on February 23, 2018; Grades and issues reported
here are found on February 23, 2018. This is a free online SSL certificate
checker.



\bibitem{vulnarebility1}https://geekflare.com/online-scan-website-security-vulnerabilities/\#1-Scan-My-Server

\bibitem{vulnarebility2}https://www.acunetix.com/online-vulnerability-scanner/

\bibitem{NIST1-2011}E. Barker and A. Roginsky, Transitions:
Recommendation for transitioning the use of cryptographic algorithms
and key lengths, NIST Special Publication 800-131A
(2011); available at http://www.gocs.eu/pages/fachberichte/archiv/075-sp800-131A.pdf


\bibitem{NIST2-2017} E. Barker,  \textquotedbl{}SP 800-67 Rev.
2, Recommendation for Triple Data Encryption Algorithm (TDEA) Block
Cipher.\textquotedbl{} NIST special publication \textbf{800} (2017) 67.

\bibitem{forENISA1}W. K. Hon and C. Millard, Banking in the cloud: Part 1-banks' use of cloud services, Computer Law and Security Review \textbf{34} (2018) 4.

\bibitem{TLS13}https://www.bleepingcomputer.com/news/security/ietf-approves-tls-13-as-internet-standard/

\bibitem{post-qua}D. J. Bernstein and T. Lange,   Post-quantum
cryptography,  Nature, \textbf{549} (2017) 188.

\bibitem{qrng}M. Herrero-Collantes and J. C. Garcia-Escartin, Quantum random number generators, Rev.
of Mod. Phys. \textbf{89} (2017) 015004.

\bibitem{SQ}C. Shukla, K. Thapliyal and A. Pathak, Semi-quantum
communication: Protocols for key agreement, controlled secure direct
communication and dialogue, Quantum Inf. Process. \textbf{16}  (2017) 295.

\bibitem{QPC}K. Thapliyal, R. D. Sharma, and A. Pathak, Orthogonal-state-based
and semi-quantum protocols for quantum private comparison in noisy
environment, arxiv: 1608.00101v1 (2016).

\bibitem{QC}N. Gisin, G. Ribordy, W. Tittel and H. Zbinden,
Quantum cryptography, Rev. Mod. Phys. \textbf{74} (2002) 145.

\bibitem{lo}H.-K. Lo, M. Curty and K. Tamaki, Secure quantum key distribution Nature Photonics \textbf{8}
(2014) 595.


\end{thebibliography}
\end{document}